\documentclass[12pt]{revtex4}
\usepackage{amsfonts}
\usepackage{amsmath}
\usepackage{amssymb}
\usepackage{graphicx}
\usepackage{color}

\begin{document}

\title{Navier-Stokes, Gross-Pitaevskii and Generalized Diffusion Equations
using Stochastic Variational Method}
\author{T. Koide}
\author{T. Kodama}
\affiliation{Instituto de F\'{\i}sica, Universidade Federal do Rio de Janeiro, C.P.
68528, 21941-972, Rio de Janeiro, Brazil}

\begin{abstract}
The stochastic variational method is applied to particle systems and
continuum mediums. As the brief review of this method, we first discuss the
application to particle Lagrangians and derive a diffusion-type equation and
the Schr\"{o}dinger equation with the minimum gauge coupling. We further
extend the application of the stochastic variational method to Lagrangians
of continuum mediums and show that the Navier-Stokes, Gross-Pitaevskii and
generalized diffusion equations are derived. The correction term for the
Navier-Stokes equation is also obtained in this method. We discuss the
meaning of this correction by comparing with the diffusion equation.
\end{abstract}

\maketitle

\section{Introduction}

The variational principle plays fundamental roles in many branches in
physics, such as analytic mechanics, field theory, statistical mechanics,
quantum many-body problems, scattering theory, fluid dynamics and so on. Due
to its elegant mathematical nature, the approach serves also as
important guidance to formulate models, extracting the essential physical
structure of the system in question. In particular, the method is
indispensable when we deal with symmetries and the associated conservation
laws of a system \cite{lancz}. All the consequences of symmetries, at least
in the classical level, are simply condensed in the fact that the Lagrangian
(density) is invariant under the transformations of the corresponding
symmetry.

On the other hand, once some dissipative processes are involved, the usual
variational approach looses its elegance. To include dissipative effects,
one has to introduce some extra function, such as the so-called Rayleigh
dissipation function or a time dependent external factor for a Lagrangian 
\cite{ray}. The choice of the functions are usually ambiguous. This
modification is attributed to the fact that dissipation is essentially
related to the energy conversion from macroscopic degrees of freedom to
internal microscopic motions, accompanying with the entropy production \cite%
{lancz}. In this sense, dissipation is beyond the scope of the classical
variational method \cite{ff,becker}.

Around third decades ago, a new variational method was proposed \cite%
{yasue1,yasueother}, where dynamical variables are extended to stochastic
quantities. In this stochastic variational method (SVM), the effects of the
microscopic degrees of freedom to the macroscopic dynamical variables are
represented in terms of noises, leading naturally to dissipation of the
dynamical variables. The advantage of this method is that we do not need to
modify Lagrangians by introducing extra functions, and, once the noise is
specified for a given Lagrangian, the dissipative dynamics is derived in a
straightforward way.

Since then, SVM has been studied for several cases, exclusively for particle
Lagrangians. It was shown that diffusion-type equations \cite%
{hasegawa,quantum_dif,davi1} and the Schr\"{o}dinger equation \cite%
{zam,gross,sch+gauge,davi2} can be derived in the framework of SVM. In
particular, in the latter case, SVM approach has been extensively discussed
in relation to Schr\"odinger and Nelson's stochastic formulation of quantum
mechanics \cite{sch,nelson}.

On the other hand, it seems that, in spite of the interesting aspects of the
method, the potential possibility of SVM has not been explored sufficiently.
Recently, we have shown that the method can be used to derive the
Navier-Stokes equation for compressible fluids \cite{kk}. In the present
work, we follow up the previous report, and develop the discussion of SVM in
more general way. Starting from a short summary of the method for the case
of particle Lagrangians, we extend the discussion to the case of continuum
mediums. We show the detailed derivations of the Navier-Stokes,
Gross-Pitaevskii and generalized diffusion equations in the framework of SVM.

Moreover, SVM can calculate a correction term to the Navier-Stokes equation.
The physical meaning of the correction is discussed by comparing with the
usual diffusion equation.

This paper is organized as follows. In Sec. II, we summarize the classical
variational method to be contrasted to SVM. In Sec. III, the application of
SVM to particle systems of four different examples. One of the results
obtained in Sec. III is equivalent to the Schr\"{o}dinger equation. The same
argument is applicable even when there is the gauge interaction as is
discussed in Sec. IV. There the gauge fields are treat as the external
fields. In Sec. V, SVM is applied to continuum mediums, and the
Navier-Stokes, Gross-Pitaevskii and generalized diffusion equations are
derived. The possible generalization of the Navier-Stokes equation is
further discussed by comparing with the diffusion equation. Section VI is
devoted to concluding remarks.

\section{Classical Variational Method}

To clarify the difference between the classical variational method and SVM,
let us restate the classical variational method in the following form.
Suppose that there exists a vector field $\mathbf{f}=\mathbf{f}(\mathbf{r},t)
$ where the velocity of a particle at the position $\mathbf{x}(t)$ is given
by 
\begin{equation}
\frac{d}{dt}\mathbf{x}(t)=\left. \mathbf{f}(\mathbf{r},t)\right\vert _{%
\mathbf{r}=\mathbf{x}(t)}.  \label{dete_eq}
\end{equation}%
As is shown, 
this quantity coincides with the particle velocity at the position of the particle 
$\mathbf{r}=\mathbf{x}(t)$, but is defined for all ${\bf r}$ at a given $t$.
This is essential when we introduce the stochastic nature for the particle
trajectory. Our strategy is to determine this field $\mathbf{f}$ by the
variational principle.

Let us consider the usual action for a particle, which is given by the
subtraction of the kinetic term and the potential term, 
\begin{equation}
I=\int_{t_{a}}^{t_{b}}dt\left[ \frac{m}{2}\left( \frac{d\mathbf{x}(t)}{dt}%
\right) ^{2}-V(\mathbf{x}(t))\right],  \label{class_act}
\end{equation}%
where $m$ is the mass of a particle and $V$ is the potential energy.

Let us consider the variation of the position, $\mathbf{x}\longrightarrow 
\mathbf{x}+\delta \mathbf{x}$, with the conditions $\delta \mathbf{x}\left(
t_{a}\right) =\delta \mathbf{x}\left( t_{b}\right) =0$ as usual. The
variation of the action is then given by 
\begin{eqnarray}
\delta I &=&\int_{t_{a}}^{t_{b}}dt\left[ m\frac{d\mathbf{x}(t)}{dt}\delta
\left( \frac{d\mathbf{x}(t)}{dt}\right) -\delta V(\mathbf{x}(t))\right] 
\notag \\
&=&\int_{t_{a}}^{t_{b}}dt\left[ -m\frac{d}{dt}\mathbf{f}(\mathbf{x}(t),t)-%
\mathbf{\nabla }V(\mathbf{x}(t))\right] \cdot {\delta \mathbf{x}},
\label{stationary_cla}
\end{eqnarray}%
and the stationarity condition of the action leads to%
\begin{equation}
m(\partial _{t}+\mathbf{f}(\mathbf{x}(t),t)\cdot \mathbf{\nabla })\mathbf{f}(%
\mathbf{x}(t),t)+\mathbf{\nabla }\ V(\mathbf{x}(t))=0.  \label{feq}
\end{equation}%
Of course, if we introduce the following variable, 
\begin{equation}
\mathbf{v}(t)=\mathbf{f}(\mathbf{x}(t),t),  \label{Vel}
\end{equation}%
then Eq. (\ref{feq}), together with the definition of the particle velocity (%
\ref{dete_eq}), can be casted into the usual form of Newton's equation of
motion,
\begin{subequations}
\begin{eqnarray}
&&\frac{d\mathbf{x}(t)}{dt}=\mathbf{v}(t), \\
&&m\frac{d\mathbf{v}(t)}{dt}+\nabla V(\mathbf{x}(t))=0,
\end{eqnarray}
\end{subequations}
where we used the definition of the material derivative, 
\begin{equation}
\frac{d\mathbf{v}(t)}{dt}=\frac{d\mathbf{f}(\mathbf{x}(t),t)}{dt}=(\partial
_{t}+\mathbf{f}(\mathbf{x}(t),t)\cdot \nabla )\mathbf{f}(\mathbf{x}(t),t).
\end{equation}

On the other hand, we can regard Eq.(\ref{Vel}) as a special value of the
solution of a partial differential equation for the field $\mathbf{f}(%
\mathbf{r},t)$,%
\begin{equation}
m(\partial _{t}+\mathbf{f}(\mathbf{r},t)\cdot \nabla )\mathbf{f}(\mathbf{r}%
,t)+\nabla V(\mathbf{r})=0,  \label{fpartial}
\end{equation}%
with some appropriate initial and boundary conditions. In the case of the
classical variational principle, we do not need to calculate $\mathbf{f}(%
\mathbf{r},t)$ for all $\mathbf{r}$ at a given $t$ to determine the
classical particle trajectory. However, when we extend $\mathbf{x}(t)$ to a
stochastic quantity, as we will see soon later, we need to know the values
of $\mathbf{f}(\mathbf{r},t)$ not only for the trajectory of a particle, but
also for general values of $\mathbf{r}$. This situation remind us the
path-integral formulation of quantum mechanics where the knowledge of action
is required as for general values of $\mathbf{r}$.

\section{stochastic variational method for particle systems}

In the classical variational method, the evolution of a particle is
deterministic. In SVM, we consider the case where the trajectory of a
particle is stochastic \cite{yasue1}. This is because we would like to
approximately take into account the effects of some microscopic degrees of
freedom, which do not appear as the dynamical variables in a given action,
through the effect of noises. Therefore, instead of Eq. (\ref{dete_eq}), the
time evolution of the particle is described by the following stochastic
differential equation (SDE), 
\begin{equation}
d\mathbf{x}(t)=\mathbf{f}(\mathbf{x}(t),t)dt+\sqrt{2\nu }d\mathbf{W}%
_{t}~~~(dt>0),  \label{fsde}
\end{equation}%
where the last term represents the noise. Properties of the noise will
depend on the detailed nature of the underlying microscopic dynamics.
However, if there are clear separation of microscopic and macroscopic time
scales, we can assume that microscopic information is approximately replaced
by the Gaussian white noise as the most standard and simplest choice. Thus, $%
d\mathbf{W}_{t} $ is differential of a Wiener process, satisfying 
\begin{subequations}
\begin{eqnarray}
E[d\mathbf{W}_{t}] &=&0, \\
E[dW_{t}^{i}~dW_{t}^{j}] &=&\delta ^{ij}dt.
\end{eqnarray}
\label{noise1}
\end{subequations}
where $E[~~~]$ denotes the expectation value for the stochastic processes.
The magnitude of the noise is characterized by $\nu$ which is a constant. We
call Eq.(\ref{fsde}) the forward SDE, because it is defined only for $dt>0$
(see the discussion below).

In stochastic systems, the particle distribution function is introduced as 
\begin{equation}
\rho (\mathbf{r},t)=E[\delta (\mathbf{r}-\mathbf{x}(t))],  \label{rho}
\end{equation}%
where the expectation value is taken over all the trajectories $\left\{ 
\mathbf{x}(t)\right\} .$ Therefore, $\rho (\mathbf{r},t)$ contains the
information of the time evolution for a given initial distribution $\rho (%
\mathbf{r},t_{0}).$ Of course, the distribution of initial values $\left\{ 
\mathbf{x}(t_{0})\right\} $ for Eq. (\ref{fsde}) should be given by $\rho (%
\mathbf{r},t_{0})$ so that Eq.(\ref{rho}) is valid at $t=t_{0}$. From this
definition and using the properties of the noise, the evolution equation of $%
\rho $ is expressed in terms of the following Fokker-Plank equation \cite%
{handbook}, 
\begin{equation}
\partial _{t}\rho (\mathbf{r},t)=\nabla \cdot (-\mathbf{f}(\mathbf{r},t)+\nu
\nabla )\rho (\mathbf{r},t).  \label{ffp}
\end{equation}

As discussed in the previous section, the field $\mathbf{f}(\mathbf{r},t)$
is yet unknown and is determined by employing the variational principle.
However, differently from the classical deterministic case, we now face to a
new problem: the trajectories of stochastic variables are not smooth and are
non-differentiable. Thus the time derivative is not uniquely defined. For
example, we may consider the following two possible definitions for the
velocity of the particle at $t$, 
\begin{subequations}
\begin{eqnarray}
\mathbf{v}(t) &=&\lim_{dt\rightarrow 0+}\frac{\mathbf{x}(t+dt)-\mathbf{x}(t)%
}{dt},  \label{vforward} \\
\tilde{\mathbf{v}}(t) &=&\lim_{dt\rightarrow 0-}\frac{\mathbf{x}(t+dt)-%
\mathbf{x}(t)}{dt}.  \label{vbackward}
\end{eqnarray}
\end{subequations}
Obviously when the particle trajectory $\mathbf{x}(t)$ is continuous and
smooth, the two definitions coincide, $\mathbf{v}(t)=\tilde{\mathbf{v}}(t)$,
as is the case of the classical variational method. On the other hand, for
the stochastic trajectory, $\mathbf{v}(t)\neq \tilde{\mathbf{v}}(t)$ in
general. In spite of this, it was shown that the variational approach can be
consistently formulated by completing the information of the backward SDE in
addition to the forward SDE, as is shown in Ref. \cite{yasue1} \footnote{%
The stochastic process of this kind is known as the Bernstein process in
Mathematics. The Bernstein process can be one of the Markov processes \cite%
{ber}.}.

Let us introduce a SDE for the time reversed evolution of $\mathbf{x}(t)$ as 
\begin{equation}
d\mathbf{x}(t)=\tilde{\mathbf{f}}(\mathbf{x}(t),t)dt+\sqrt{2\nu }d\tilde{%
\mathbf{W}}_{t}~~~(dt<0),  \label{Backward}
\end{equation}%
where we assume the properties of the new noises $d\tilde{\mathbf{W}}_{t}\ $%
stay the same as before, 
\begin{subequations}
\begin{eqnarray}
E[d\tilde{\mathbf{W}}_{t}] &=&0, \\
E[d\tilde{W}_{t}^{i}~d\tilde{W}_{t}^{j}] &=&\delta ^{ij}|dt|.
\end{eqnarray}
\label{noise2}
\end{subequations}
We call Eq.(\ref{Backward}) the backward SDE. We can obtain the Fokker-Plank
equation from this backward SDE similarly to the case of the forward SDE as 
\begin{equation}
\partial _{t}\rho (\mathbf{r},t)=\nabla \cdot (-\tilde{\mathbf{f}}(\mathbf{r}%
,t)-\nu \nabla )\rho (\mathbf{r},t).  \label{bfp}
\end{equation}%
If the above backward SDE, as a set of stochastic processes, describes the
correct representation of the time-reversed process of the forward SDE, the
Fokker-Plank equation (\ref{bfp}) should coincide with Eq. (\ref{ffp}). From
this condition, we obtain 
\begin{equation}
\mathbf{f}(\mathbf{r},t)=\tilde{\mathbf{f}}(\mathbf{r},t)+2\nu \nabla \ln
\rho (\mathbf{r},t)+\nabla \times \mathbf{A}(\mathbf{r},t)+\mathbf{B}(t),
\end{equation}%
were $\mathbf{A}(\mathbf{r},t)$ and $\mathbf{B}(t)$ are arbitrary functions.
In the absence of external forces and non-singular boundary conditions, we
can simply omit $\mathbf{A}(\mathbf{r},t)$ and $\mathbf{B}(t)$, so that $%
\mathbf{f}(\mathbf{r},t)$ and $\tilde{\mathbf{f}}(\mathbf{r},t)$ are related
as 
\begin{equation}
\mathbf{f}(\mathbf{r},t)=\tilde{\mathbf{f}}(\mathbf{r},t)+2\nu \nabla \ln
\rho (\mathbf{r},t).  \label{consis}
\end{equation}

In order to incorporate the forward and backward SDEs consistently in the
variational scheme of stochastic variables \cite{nelson,yasue1}, we
introduce the mean forward derivative, 
\begin{equation}
D\mathbf{x}(t)\equiv \lim_{h\rightarrow 0+}E\left[ \frac{\mathbf{x}(t+h)-%
\mathbf{x}(t)}{h}\Big{|}\mathcal{P}_{t}\right] ,  \label{Dx}
\end{equation}%
and the mean backward derivative, 
\begin{equation}
\tilde{D}\mathbf{x}(t)\equiv \lim_{h\rightarrow 0+}E\left[ \frac{\mathbf{x}%
(t)-\mathbf{x}(t-h)}{h}\Big{|}\mathcal{F}_{t}\right] ,  \label{DxTilda}
\end{equation}%
instead of the two velocities, Eqs. (\ref{vforward}) and (\ref{vbackward}).
Here, $E[F(t^{\prime })|\mathcal{P}_{t}]$ denotes the conditional average of
a time sequence of stochastic variables $\left\{ F\left( t^{\prime }\right)
,t_{a}<t^{\prime }<t_{b}\right\} $ only for $t^{\prime }>t$, fixing the
values of $F\left( t^{\prime }\right) $ for $t\prime \leq t$ \cite{nelson}.
Thus, $DF(t)$ is still a stochastic quantity, defined by the forward SDE. In
fact, this definition leads to 
\begin{equation}
D\mathbf{x}(t)=\mathbf{f}(\mathbf{x}(t),t).  \label{Dx=f}
\end{equation}%
Similarly, $E[F(t^{\prime })|\mathcal{F}_{t}]$ is the conditional average
for the past sequence, defined by the backward SDE, fixing the future
values, so that 
\begin{equation}
\tilde{D}\mathbf{x}(t)=\tilde{\mathbf{f}}(\mathbf{x}(t),t).
\label{Dxtilda=ftilda}
\end{equation}
These velocities are used to replace the classical velocity in the kinetic
term of the action. However, this replacement is not unique and each case
corresponds to different physical scenarios. In the following, we discuss
four different cases.

\subsection{CASE 1: Pure Forward Derivative}

Let us consider the case where the velocity $d\mathbf{x}/dt$ in the
classical action is replaced only by the mean forward derivative $D\mathbf{x}%
(t)$. Then the action is given by 
\begin{equation}
I=\int_{t_a}^{t_b}dtE\left[ \frac{m}{2} D\mathbf{x}(t) \cdot D\mathbf{x}(t)
-V(\mathbf{x}(t))\right].  \label{pact_1}
\end{equation}
Note that the products of the mean forward (backward) derivatives are
independent of the choice of the discretization scheme such as the Ito,
Stratonovich-Fisk and H\"anggi-Klimontovich schemes by the definition of the
mean forward (or backward) derivative \cite{handbook}.

Now we introduce the variation of the position of the particle as 
\begin{equation}
\mathbf{x}(t)\longrightarrow \mathbf{x}_{\lambda }(t)=\mathbf{x}(t)+\lambda 
\mathbf{g}(\mathbf{x}(t),t),
\end{equation}%
where $\lambda $ is an expansion parameter and $\mathbf{g}(\mathbf{r},t)$ is
an arbitrary smooth function which is differentiable and satisfies the
boundary condition, 
\begin{equation}
\mathbf{g}(\mathbf{x}(t_{a,b}),t_{a,b})=0.
\end{equation}%
Substituting into the action (\ref{pact_1}), we obtain 
\begin{eqnarray}
I_{\lambda } &=&I+\lambda \int_{t_a}^{t_b}dtE\left[ m D\mathbf{x}(t) \cdot D%
\mathbf{g}(\mathbf{x}(t),t) -\nabla V(\mathbf{r})|_{\mathbf{r}=\mathbf{x}%
(t)}\cdot \mathbf{g}(\mathbf{x}(t),t)\right] +O(\lambda ^{2})  \notag \\
&=&I+\lambda \int_{t_a}^{t_b}dtE\left[ \{-m\tilde{D}(D\mathbf{x}(t))-\nabla
V(\mathbf{r})|_{\mathbf{r}=\mathbf{x}(t)}\}\cdot \mathbf{g}(\mathbf{x}(t),t)%
\right] +O(\lambda ^{2})  \notag \\
&=&I+\lambda \int_{t_a}^{t_b}dtE\left[ \{-m\tilde{D}\mathbf{f}(\mathbf{x}%
(t),t)-\nabla V(\mathbf{r})|_{\mathbf{r}=\mathbf{x}(t)}\}\cdot \mathbf{g}(%
\mathbf{x}(t),t)\right] +O(\lambda ^{2}).
\end{eqnarray}%
Here, we have used the stochastic partial integral formula \cite{zm,kk} and
Eq. (\ref{Dx=f}). Using the Ito formula (Ito's lemma) \cite{handbook}, the
first term in the integral can be calculated as%
\begin{equation}
\tilde{D}\mathbf{f}(\mathbf{x}(t),t)=\left. \left( \partial _{t}+\tilde{%
\mathbf{f}}(\mathbf{r},t)\cdot \nabla -\nu \nabla ^{2}\right) \mathbf{f}(%
\mathbf{r},t)\right\vert _{\mathbf{r}=\mathbf{x}(t)}.
\end{equation}%
Therefore, the stationary condition of the action under the variations leads
to 
\begin{equation}
m[\partial _{t}+\mathbf{f}(\mathbf{r},t)\cdot \nabla -2\nu (\nabla \ln \rho (%
\mathbf{r},t))\cdot \nabla -\nu \nabla ^{2}]~\mathbf{f}(\mathbf{r}%
,t)=-\nabla V(\mathbf{r}).  \label{pf1}
\end{equation}%
Here $\tilde{\mathbf{f}}(\mathbf{r},t)$ is replaced by $\mathbf{f}(\mathbf{r}%
,t)$ using the consistency condition (\ref{consis}). Equation (\ref{pf1})
and the Fokker-Planck equation (\ref{ffp}) constitute a system of partial
differential equations for the unknown function $\mathbf{f}(\mathbf{r},t)$
and $\rho (\mathbf{r},t)$.

As stressed in the previous section, in the classical variational method,
the final equation of motion requires only the velocity $\mathbf{v}(t)=%
\mathbf{f}(\mathbf{x}(t),t)$ and the knowledge of the field $\mathbf{f}(%
\mathbf{r},t)$ at arbitrary position $\mathbf{r}$ is not necessary. In SVM,
we, however, have to calculate the field $\mathbf{f}(\mathbf{r},t)$ itself.
In fact, $\mathbf{f}(\mathbf{x}(t),t)$ which appears in Eq. (\ref{pf1})
cannot be completely replaced by $\mathbf{v}(t)$ because of the second
spatial derivative term, which did not appear in the classical variational
method. That is, the fundamental variable is not $\mathbf{v}(t)$ but $%
\mathbf{f}(\mathbf{r},t)$ in SVM. The velocity on the trajectory is known
only after solving the forward SDE (\ref{fsde}) with the solution of Eq. (%
\ref{pf1}).

From Eq.(\ref{Dx=f}), $\mathbf{f}(\mathbf{r},t)$ gives the mean forward
velocity of the particle, obeying the forward SDE equation. Thus, one might
expect that $\mathbf{f}(\mathbf{r},t)$ is the velocity field corresponding
to the probability current. From the Fokker-Planck equation (\ref{ffp}), we
identify such a probability current as 
\begin{equation}
\mathbf{J}(\mathbf{r},t)=(\mathbf{f}(\mathbf{r},t)-\nu \nabla )\rho (\mathbf{%
r},t).
\end{equation}
One can see that $\mathbf{f}(\mathbf{r},t)$ is not the velocity field
parallel to the probability current, and neither $\tilde{\mathbf{f}}(\mathbf{%
r},t)$ is. However, the average of $\mathbf{f}(\mathbf{r},t)$ and $\tilde{%
\mathbf{f}}(\mathbf{r},t),$ 
\begin{equation}
\mathbf{f}_{m}(\mathbf{r},t)=\frac{\mathbf{f}(\mathbf{r},t)+\tilde{\mathbf{f}%
}(\mathbf{r},t)}{2}=\mathbf{f}(\mathbf{r},t)-\nu \nabla \ln \rho(\mathbf{r}%
,t) =\tilde{\mathbf{f}}(\mathbf{r},t)+\nu \nabla \ln \rho (\mathbf{r},t),
\label{vel_ave}
\end{equation}%
is exactly related to the probability current as 
\begin{equation}
\mathbf{J}(\mathbf{r},t)=\rho (\mathbf{r},t)\mathbf{f}_{m}(\mathbf{r},t).
\end{equation}%
That is, the velocity parallel to $\mathbf{J}(\mathbf{r},t)$ is given by $%
\mathbf{f}_m(\mathbf{r},t)$.

Equation (\ref{pf1}) is then re-expressed as 
\begin{equation}
\rho (\partial _{t}+\mathbf{f}_{m}\cdot \nabla )\ f_{m}^{\ i}-\nabla \mathbf{%
\cdot }(\nu \rho \partial _{i}\ \mathbf{f}_{m})-\nabla \mathbf{\cdot }(\nu
\rho \mathbf{\nabla }\ f_{m}^{\ i})-\nabla \mathbf{\cdot }[(\nu \rho )\nabla 
\mathbf{\ }(\nu \ \partial _{i}\ln \rho )]=-\frac{\rho }{m}\nabla^i V,
\label{pf1_2}
\end{equation}
where $f_{m}^{\ i}$ is the $i$-th component of $\mathbf{f}_{m}.$ In terms of
the mean velocity field, $\mathbf{f}_{m},$ the Fokker-Planck equation for $%
\rho $ is reduced simply to the continuity equation,%
\begin{equation}
\frac{\partial \rho }{\partial t}+\nabla \cdot \left( \rho \mathbf{f}%
_{m}\right) =0.  \label{Continuity}
\end{equation}%
Equations (\ref{pf1_2}) and (\ref{Continuity}) constitute a closed set of
differential equations for $\mathbf{f}_{m}(\mathbf{r},t)$ and $\rho (\mathbf{%
r},t)$. Note that Eq.(\ref{pf1_2}) contains the term which violates the time
reversal symmetry and describes a dissipative process.

\subsection{ CASE 2 : Pure Backward Derivative}

Another possibility is to use only the mean backward derivative to express
the kinetic term, 
\begin{equation}
I=\int_{t_a}^{t_b}dtE\left[ \frac{m}{2} \tilde{D}\mathbf{x}(t) \cdot \tilde{D%
}\mathbf{x}(t) -V(\mathbf{x}(t))\right] .  \label{pact_2}
\end{equation}%
Repeating the variational procedure analogous to the case 1, we obtain%
\begin{equation}
I_{\lambda }=I+\lambda \int_{t_a}^{t_b}dtE\left[ \{-m{D}\tilde{\mathbf{f}}(%
\mathbf{x}(t),t)-\nabla V(\mathbf{r})|_{\mathbf{r}=\mathbf{x}(t)}\}\cdot 
\mathbf{g}(\mathbf{x}(t),t)\right] +O(\lambda ^{2}).
\end{equation}%
The stationarity condition of the action leads to, instead of Eq.(\ref{pf1}%
), 
\begin{equation}
m[\partial _{t}+\tilde{\mathbf{f}}(\mathbf{r},t)\cdot \nabla +2\nu (\nabla
\ln \rho (\mathbf{r},t))\cdot \nabla +\nu \nabla ^{2}]\tilde{\mathbf{f}}(%
\mathbf{r},t)=-\nabla V(\mathbf{r}).  \label{pf2}
\end{equation}%
Equation for $\mathbf{f}_{m}$ becomes 
\begin{equation}
\rho (\partial _{t}+\mathbf{f}_{m}\cdot \nabla )\ f_{m}^{\ i}+\nabla \mathbf{%
\cdot }(\nu \rho \partial _{i}\ \mathbf{f}_{m})+\nabla \mathbf{\cdot }(\nu
\rho \mathbf{\nabla }\ f_{m}^{\ i})-\nabla \mathbf{\cdot }[(\nu \rho )\nabla 
\mathbf{\ }(\nu \ \partial _{i}\ln \rho )]=-\frac{\rho }{m}\nabla^i V.
\label{pf2_2}
\end{equation}%
One can see that this equation describes the time reversed process of Eq. (%
\ref{pf1_2}). In fact, Eq. (\ref{pf2_2}) coincides with Eq. (\ref{pf1_2}) by
exchanging the sign of $\nu $.

\subsection{CASE 3: Mixed Product of Two Derivatives}

Hasegawa \cite{hasegawa} studied the case where the kinetic term is written
with both of $D\mathbf{x}$ and $\tilde{D}\mathbf{x}$ as, 
\begin{equation}
I=\int_{t_a}^{t_b}dtE\left[ \frac{m}{2} {D}\mathbf{x}(t) \cdot \tilde{D}%
\mathbf{x}(t) -V(\mathbf{x}(t))\right] .  \label{pact_3}
\end{equation}%
Differently from the previous two cases, this kinetic term is symmetric for
the exchange between $D$ and $\tilde{D}$. The stochastic variation, then,
leads to 
\begin{equation}
(\partial _{t}+\mathbf{f}_{m}\cdot \nabla ) f^i_{m}+2\nu ^{2}\nabla^i (\rho
^{-1/2}\nabla ^{2}\sqrt{\rho })=-\frac{1}{m}\nabla^i V.  \label{case3_vm}
\end{equation}%
Differently from Eqs. (\ref{pf1_2}) and (\ref{pf2_2}), one can see that this
equation is symmetric for the time reversed operation $t\leftrightarrow -t$.
This is the consequence of the symmetric replacement in the kinetic term.

One can confirm that Eq. (\ref{pact_3}) is completely equivalent to the
result of Ref. \cite{hasegawa}, by noting the relation $\mathbf{f}_{m}(%
\mathbf{r},t)=2\nu \mathbf{\nabla }S(\mathbf{r},t)$.

\subsection{CASE 4: Average of Cases 1 and 2 - Schr\"{o}dinger Equation}

\label{sec:ave}

As the last example, we consider that the kinetic term of the action is
given by the average of the mean forward and backward derivatives \cite{zam}%
, 
\begin{equation}
I=\int_{t_a}^{t_b}dtE\left[ \frac{m}{2}\frac{D\mathbf{x}(t) \cdot D\mathbf{x}%
(t) + \tilde{D}\mathbf{x}(t)\cdot \tilde{D}\mathbf{x}(t)}{2}-V(\mathbf{x}(t))%
\right] .  \label{pact_4}
\end{equation}%
In this case, the kinetic term is symmetric for the exchange of $D$ and $%
\tilde{D}$, and hence the derived equation by the variation is symmetric for
the time reversed operation. As a matter of fact, SVM leads to the following
equation, 
\begin{equation}
\frac{m}{2}\left\{ \partial _{t}\tilde{\mathbf{f}}+\left( \mathbf{f}\cdot
\nabla \right) \ \tilde{\mathbf{f}}+\nu \nabla ^{2}\tilde{\ \mathbf{f}}%
+\partial _{t}\mathbf{f} + ( \tilde{\mathbf{f}}\cdot \nabla ) \mathbf{f}-\nu
\nabla ^{2}\mathbf{f}\right\} =-\nabla V.
\end{equation}%
This equation can be re-expressed as 
\begin{equation}
(\partial _{t}+\mathbf{f}_{m}\cdot \nabla ) f^i_{m}-2\nu ^{2}\nabla^i (\rho
^{-1/2}\nabla ^{2}\sqrt{\rho })=-\frac{1}{m}\nabla^i V.  \label{case4_vm}
\end{equation}

It is known that Eq. (\ref{case4_vm}), together with the continuity equation
(\ref{Continuity}), is (almost) equivalent to the Schr\"{o}dinger equation 
\cite{zam}. To show this, we introduce the following real function $\theta$, 
\begin{equation}
\nabla \theta =\frac{1}{2\nu }\mathbf{f}_{m}.  \label{phase1}
\end{equation}%
Obviously, here we assumed that $\mathbf{f}_m$ is irrotational, $\nabla\cdot 
\mathbf{f}_m =0$. Then, from Eq.(\ref{case4_vm}), we obtain, 
\begin{equation}
\nabla \left[ \partial _{t}\theta +\nu \left( \nabla \theta \right) ^{2}-\nu
(\rho ^{-1/2}\nabla ^{2}\sqrt{\rho })+\frac{1}{2\nu m}V\right] =0.
\label{EqVel}
\end{equation}
This leads to 
\begin{equation}
\partial _{t}\theta =-\nu \left( \nabla \theta \right) ^{2}+\nu (\rho
^{-1/2}\nabla ^{2}\sqrt{\rho })-\frac{1}{2\nu m}V + c(t),  \label{thetadot}
\end{equation}
where $c(t)$ is an arbitrary function of $t$, However, this is always
absorbed into the definition of $\theta$. Thus without loss of generality,
we can set $c(t)=0$. On the other hand, the Fokker-Plank equation is reduced
to 
\begin{equation}
\partial _{t}\ln \sqrt{\rho }=-2\nu (\nabla \ln \sqrt{\rho })\cdot \nabla
\theta -\nu \nabla ^{2}\theta .  \label{rhodot}
\end{equation}

Now we introduce the wave function as 
\begin{equation}
\psi =\sqrt{\rho }e^{i\theta }.  \label{Psi}
\end{equation}%
Then, from Eqs.(\ref{thetadot}) and (\ref{rhodot}), the evolution equation
of $\psi$ is given by, 
\begin{equation}
i\partial _{t}\psi =\nu \left[ -\nabla ^{2}+\frac{1}{2\nu m}V\right] \psi .
\label{sch_SVM}
\end{equation}%
By choosing $\nu =\hbar /(2m)$, we obtain the Schr\"{o}dinger equation, 
\begin{equation}
i\hbar \partial _{t}\psi =\left[ \frac{1}{2m}\left( \frac{\hbar }{i}\nabla
\right) ^{2}+V\right] \psi .  \label{Sch_1}
\end{equation}%
Note that, by the definition of $\psi$ given by Eq. (\ref{Psi}), $|\psi |^{2}
$ gives naturally the probability density of the particle.

Of course, if the flow is rotational, we cannot introduce the phase of the
wave function as Eq. (\ref{phase1}). Thus the equation obtained in SVM is
not completely equivalent to the Schr\"odinger equation.

The above derivation of the Schr\"{o}dinger Equation in terms of SVM opens
an interesting possibility for a new formulation of quantum mechanics, and
extensive works have been done so far \cite{zam,davi1}, but in this present
paper, we do not pursue further this direction. We only mention that there
is an ambiguity for the choice of Lagrangian in the derivation of the Schr%
\"{o}dinger equation. Let us consider the following action 
\begin{eqnarray}
I &=&\int_{b}^{a}dt\left[ \frac{m}{2}\left( \frac{D\mathbf{x}(t) \cdot D%
\mathbf{x}(t) + \tilde{D}\mathbf{x}(t) \cdot \tilde{D}\mathbf{x}(t)}{2}-%
\frac{\alpha }{8}(D\mathbf{x}(t)-\tilde{D}\mathbf{x}(t))\cdot (D\mathbf{x}%
(t)-\tilde{D}\mathbf{x}(t))\right) \right.  \notag \\
&&\left. -V(\mathbf{x}(t))\frac{{}}{{}}\right] ,
\end{eqnarray}%
where $\alpha $ is a constant. This action is reduced to the previous action
(\ref{pact_4}) in the vanishing $\alpha $ limit. As is shown in Ref. \cite%
{davi2}, the stochastic variation of this action again leads to the Schr\"{o}%
dinger equation by choosing 
\begin{equation}
\nu =\frac{1}{\sqrt{1-\alpha /2}}\frac{\hbar }{2m}.
\end{equation}

Here, we discussed the wave function for the action given by Eq. (\ref%
{pact_4}). The wave function $\psi$ defined by Eq. (\ref{Psi}) can be
introduced even in the other cases discussed before. However, only this
fourth case leads to the known form of the Schr\"{o}dinger equation. For
example, when we apply the same argument to Eq. (\ref{case3_vm}), we obtain
(again assuming that $\mathbf{f}_{m}$ is irrotational) 
\begin{equation}
i \hbar \partial_t \psi = \left[ - \frac{\hbar^2}{2m} \nabla^2  
+ V 
\right] \psi
+ \frac{\hbar^2}{2m} 
\left[
\frac{1}{|\psi|^2} \nabla^2 |\psi|^2
- \frac{1}{2} \frac{1}{|\psi|^4} (\nabla |\psi|^2 )^2
\right] \psi .  \label{sch2}
\end{equation}
Since this equation is not linear in $\psi ,$ the superposition principle
required in quantum mechanics is not satisfied.

\section{Particle Interacting with Gauge Fields}

\label{sec_ele}

The gauge symmetry is an important principle to describe interactions among
elementary particles. Thus whether SVM can be applied consistently keeping
the gauge invariance is a crucial question. In the case of the Schr\"{o}%
dinger equation, it is known that SVM leads to the correct wave equation
with minimal coupling of a charged particle, starting from the corresponding
classical action as sketched below \cite{zam,sch+gauge}.

Let us start with the action of a classical particle interacting with the
external gauge fields, 
\begin{equation}
I=\int dt\left[ \frac{m}{2}\mathbf{v}^{2}+\frac{e}{c}\mathbf{v}\cdot \mathbf{%
A}-V-e\phi \right] ,
\end{equation}%
where $\phi $ and $\mathbf{A}$ are the scalar and vector potentials of the
electromagnetic fields, and $V$ is another potential energy, respectively.
Here we used the CGI-Gauss unit. One can easily check that this action is
invariant for the gauge transform as \footnote{%
Here, invariance means that the modification of the action by the gauge
transform is absorbed into the total time derivative term of the Lagrangian.}
\begin{subequations}
\begin{eqnarray}
\mathbf{A} &\longrightarrow &\mathbf{A}^{\prime }=\mathbf{A}+\nabla \chi , \\
\phi &\longrightarrow &\phi ^{\prime }=\phi -\frac{1}{c}\frac{\partial }{%
\partial t}\chi ,
\end{eqnarray}
\end{subequations}
where $\chi$ is an arbitrary function.

For the stochastic variations to derive the quantum mechanical wave
equation, we replace this classical action by, 
\begin{equation}
I=\int dt\left[ \frac{m}{2}\frac{D\mathbf{x}\cdot D\mathbf{x}+\tilde{D}%
\mathbf{x}\cdot \tilde{D}\mathbf{x}}{2}+\frac{e}{c}\frac{D\mathbf{x}+\tilde{D%
}\mathbf{x}}{2}\cdot \mathbf{A}-V-e\phi \right] .
\end{equation}%
The reason for this choice is that the action should be symmetric under the
time reversal. Following the standard procedure, we find that the stochastic
variation of the action becomes 
\begin{equation}
I_{\lambda }=I+\lambda \int dt\left[ -\frac{m}{2}(\tilde{D}\ \mathbf{f}+D\ 
\mathbf{\tilde{f}})-\frac{e}{2c}(D+\tilde{D})\mathbf{A}+\frac{e}{c}\sum_{i}%
\frac{f^{\ i}+\tilde{f}^{\ i}}{2}\nabla A^{i}-\nabla V-e\nabla \phi \right]
\cdot \mathbf{g}+O(\lambda ^{2}),
\end{equation}%
so that we obtain 
\begin{equation}
(\partial _{t}+\mathbf{f}_{m}\cdot \nabla )\mathbf{f}_{m}-2\nu ^{2}\nabla
(\rho ^{-1/2}\nabla ^{2}\sqrt{\rho })+\frac{1}{m}\nabla (V+e\phi )=-\frac{e}{%
cm}\partial _{t}\mathbf{A}+\frac{e}{cm}\mathbf{f}_{m}\times (\nabla \times 
\mathbf{A}).  \label{eqvm_ele}
\end{equation}%
The second term on the right hand side represents the Lorentz force.

In order to derive the corresponding Schr\"{o}dinger equation, we introduce
the phase $\theta $ by 
\begin{equation}
\nabla \theta =\frac{\mathbf{f}_{m}+\frac{e}{cm}\mathbf{A}}{2\nu },
\label{phase2}
\end{equation}%
instead of Eq. (\ref{phase1}). Then, the equation corresponding to Eq. (\ref%
{thetadot}) becomes 
\begin{equation}
\partial _{t}\theta =-\nu \left( \nabla \theta -\frac{e}{2cm\nu }\mathbf{A}%
\right) ^{2}+\nu (\rho ^{-1/2}\nabla ^{2}\sqrt{\rho })-\frac{1}{2\nu m}%
(V+e\phi ).
\end{equation}%
On the other hand, the equation of $\ln \sqrt{\rho}$ is 
\begin{equation}
\partial _{t}\ln \sqrt{\rho }=-2\nu (\nabla \ln \sqrt{\rho })\cdot (\nabla
\theta )-\nu \nabla ^{2}\theta +\frac{e}{cm}(\nabla \ln \sqrt{\rho })\cdot 
\mathbf{A}+\frac{e}{2cm}(\nabla \cdot \mathbf{A}).
\end{equation}%
Finally, we find the equation for the wave function $\psi $ defined in Eq. (%
\ref{Psi}) becomes 
\begin{equation}
i\hbar \partial _{t}\psi =\left[ \frac{1}{2m}\left( -i\hbar \nabla -\frac{e}{%
c}\mathbf{A}\right) ^{2}+V+e\phi \right] \psi ,
\end{equation}%
where again, $\nu =\hbar /(2m)$. This is exactly the Schr\"{o}dinger
equation of a charged particle, with the minimal coupling of the external
gauge fields.

Note that, in the derivation of the Schr\"{o}dinger equation, we have
assumed that the vector field $\mathbf{f}_{m}$ is irrotational. On the other
hand, in the derivation here, as is seen from the definition of the phase (%
\ref{phase2}), $\mathbf{f}_{m}$ can be rotational but $\mathbf{f}_{m}+e%
\mathbf{A}/(cm)$ should be irrotational. This fact shows that the well-known
Aharanov-Bohm effect is closely related to the rotational nature of the flow
of probability density.

\section{Continuum Medium}

So far, we have discussed the application of SVM to particle Lagrangian. In
this section, we employ to a system composed of a continuum medium \cite{kk}.

Let us consider a continuum medium described by the mass density $\rho _{m}(%
\mathbf{r},t)$ and the collective flow velocity field $\mathbf{v}(\mathbf{r}%
,t)$. Then the classical action of this system (in the Euler coordinate) is
given by a functional of the two fields, $\rho _{m}$ and $\mathbf{v}$ as 
\begin{equation}
I\left[ \rho _{m},\mathbf{v}\right] =\int_{t_b}^{t_a}dt\int d^{3}\mathbf{r}%
\left[ \frac{\rho _{m}(\mathbf{r},t)}{2}\mathbf{v}^{2}(\mathbf{r}%
,t)-\varepsilon (\rho _{m}(\mathbf{r},t))\right] ,  \label{EulerAction}
\end{equation}%
where $\varepsilon $ is the potential energy (internal energy) density of
the medium. For the sake of simplicity, we assume that the potential energy
is a function only of $\rho _{m}$. To perform the variational procedure
directly from Eq. (\ref{EulerAction}), we have to add the constraints for $%
\rho _{m}$ and $\mathbf{v}$ as is done in Refs. \cite{lancz,ff}. Due to
this, in order to introduce SVM, it is more convenient to use the Lagrangian
coordinate system $\left\{ \mathbf{R}\right\} $, rather than the Euler
coordinate system $\left\{ \mathbf{r}\right\} $ to specify the elements of
the continuum medium. As is well-known, the Lagrange coordinate system is
attached to the medium in such a way that the coordinate $\mathbf{R}$
represents the initial position of a element. In terms of this Lagrangian
coordinate system, the above action is expressed as a functional of the
trajectory of every element, 
\begin{equation}
I\left[ \mathbf{x}(\mathbf{R;}t)\right] =\int_{b}^{a}dt\int d^{3}\mathbf{R}%
\rho _{0}(\mathbf{R})\left[ \frac{1}{2}\left( \frac{d\mathbf{x}(\mathbf{R;}t)%
}{dt}\right) ^{2}-J\ \varepsilon (\rho _{m}(\mathbf{x}(\mathbf{R;}t))\right]
,  \label{ac_field}
\end{equation}%
where $\mathbf{x}(\mathbf{R;}t)$ represents the trajectory of the element
specified by $\mathbf{R,}$ $\rho _{0}(\mathbf{R})$ is the initial mass
density and $J$ is the Jacobian of the transformation,%
\begin{equation}
\mathbf{R\rightarrow r}=\mathbf{x}\left( \mathbf{R},t\right),
\end{equation}%
that is, $J=J(\mathbf{R;}t)=\mathrm{det}|\partial \mathbf{r}/\partial 
\mathbf{R}|$. Assuming that there is no chaotic flows, the conservation of
the mass of the system is expressed as 
\begin{equation}
d^{3}\mathbf{r}~\rho _{m}(\mathbf{r},t)=d^{3}\mathbf{R}~\rho _{0}(\mathbf{R}%
),
\end{equation}%
so that we have the following relation, 
\begin{equation}
\rho _{m}(\mathbf{r},t)=\frac{1}{J}\rho _{0}(\mathbf{R}).
\end{equation}%
It is easy to see that the classical variational principle for the action (%
\ref{ac_field}) with respect to the trajectories $\mathbf{x}(\mathbf{R;}t)$
leads to the Euler equation \cite{kk}.

In terms of the Lagrangian coordinate, the implementation of SVM is
straightforward. First the trajectory of the Lagrangian coordinate is
determined by the following forward and backward SDEs, 
\begin{subequations}
\begin{eqnarray}
d\mathbf{x} &=&\mathbf{f}(\mathbf{x},t)dt+\sqrt{2\nu }d\mathbf{W}%
(t),~~~(dt>0), \\
d\mathbf{x} &=&\tilde{\mathbf{f}}(\mathbf{x},t)dt+\sqrt{2\nu }d\tilde{%
\mathbf{W}}(t),~~~(dt<0),
\end{eqnarray}
\end{subequations}
where $d\mathbf{W}(t)$ and $d\tilde{\mathbf{W}}(t)$ are noise terms whose
correlation properties are same as Eqs. (\ref{noise1}) and (\ref{noise2}).
As before, we require that unknown fields $\mathbf{f}$ and $\tilde{\mathbf{f}%
}$ satisfy the consistency condition (\ref{consis}). Then the continuity
equation obtained from these Fokker-Plank equations is expressed with the
mass density as 
\begin{equation}
\frac{\partial \rho_m}{\partial t} + \nabla \cdot (\rho_m \mathbf{f}_m ) =0.
\label{Continuity2}
\end{equation}
From now on, to simplify the notation, we will omit showing explicit the $%
\left( \mathbf{R},t\right) $ dependence of the trajectories, $\mathbf{x}=%
\mathbf{x}\left( \mathbf{R},t\right) ,$ and the density, $\rho _{m}=\rho
_{m}\left( \mathbf{R},t\right) $.

Similarly to the previous section, we will replace the kinetic term with the
forward and/or backward mean derivatives. For example, we chose 
\begin{equation}
I=\int_{t_b}^{t_a}dt\int d^{3}\mathbf{R}\rho _{0}(\mathbf{R})\left[ \frac{1}{%
2} D\mathbf{x}\cdot D\mathbf{x} - J\ \varepsilon (\rho _{m})\right] .
\end{equation}
As the variation, we consider 
\begin{equation}
\mathbf{x}\left( \mathbf{R},t\right) \longrightarrow \mathbf{x}_{\lambda
}\left( \mathbf{R},t\right) =\mathbf{x}\left( \mathbf{R},t\right) +\lambda 
\mathbf{g}(\mathbf{x}\left( \mathbf{R},t\right) ,t),  \label{vari}
\end{equation}%
where $\lambda $ is an expansion parameter and $\mathbf{g}(\mathbf{r},t)$ is
an arbitrary function. Under this variation, the mass density $\rho_m$ is
changed as, 
\begin{equation}
\delta _{\lambda }\rho _{m}=\rho _{m}\left( \mathbf{r}_{\lambda },t\right)
-\rho_m (\mathbf{r},t)=-\lambda \frac{\rho_m }{J}\sum_{ij}A_{ij}\frac{%
\partial g^{i}}{\partial R^{j}}+O(\lambda ^{2}).
\end{equation}%
Here $A_{ij}$ is the cofactor $\left( ij\right) $ of the Jacobian\ $J$,
which satisfies the following properties, 
\begin{subequations}
\begin{eqnarray}
A_{im} &\equiv &\frac{\partial J}{\partial (\partial r_{i}/\partial R_{m})}=(%
\mathbf{\vec{\nabla}}r_{j}\times \mathbf{\vec{\nabla}}r_{k})_{m},\ \ 
\label{PropAij-1} \\
\sum_{j}\frac{\partial }{\partial R_{j}}A_{ij} &=&\mathbf{\vec{\nabla}}\cdot
(\mathbf{\vec{\nabla}}r_{j}\times \mathbf{\vec{\nabla}}r_{k})=0,
\label{PropAij-2} \\
\sum_{l}\frac{\partial r_{l}}{\partial R_{i}}A_{lj} &=&\sum_{l}\frac{%
\partial r_{i}}{\partial R_{l}}A_{jl}=J\delta _{ij},  \label{PropAij-3} \\
\sum_{j}A_{ij}\frac{\partial }{\partial R_{j}} &=&\sum_{j}A_{ij}\sum_{l}%
\frac{\partial r_{l}}{\partial R_{j}}\frac{\partial }{\partial r_{l}}%
=\sum_{l}J\delta _{il}\partial _{r_{l}}=J\partial _{r_{i}},
\label{PropAij-4}
\end{eqnarray}
\end{subequations}
where $\mathbf{\vec{\nabla}}$ denotes the gradient with respect to $\mathbf{R%
}$ which should be distinguished from $\nabla .$ The indices $\left(
i,j,k\right) \ \ $in Eqs. (\ref{PropAij-1}) and (\ref{PropAij-2}) should be
taken as cyclic. By using these properties, the action under the variation
is expressed as 
\begin{eqnarray}
I_{\lambda } &=&I-\lambda \int_{a}^{b}dt\int d^{3}\mathbf{R}\ \rho _{m0}E%
\left[ \lambda \mathbf{g}\cdot \left( \tilde{D}D\mathbf{x}(\mathbf{R},t)+%
\frac{1}{\rho _{m}}\nabla \left( -\frac{d}{d(1/\rho _{m})}\frac{\varepsilon 
}{\rho _{m}}\right) \right) \right] +O(\lambda ^{2}).  \notag \\
&&
\end{eqnarray}

Then the stationary condition for the variations 
\begin{equation}
\tilde{D}D\mathbf{x}(\mathbf{R},t)+\frac{1}{\rho _{m}}\nabla \left( -\frac{d%
}{d(1/\rho _{m})}\frac{\varepsilon }{\rho _{m}}\right) =0,
\end{equation}%
leads to (\textit{conf}. \ref{pf1}) 
\begin{equation}
\rho _{m}~[\partial _{t}+\mathbf{f}(\mathbf{r},t)\cdot \nabla _{r}-2\nu
(\nabla \ln \rho _{m}(\mathbf{r},t))\cdot \nabla -\nu \nabla ^{2}]~\mathbf{f}%
(\mathbf{r},t)+\nabla \left( -\frac{d}{d(1/\rho _{m})}\frac{\varepsilon }{%
\rho _{m}}\right) =0.
\end{equation}%
By using Eq. (\ref{vel_ave}), this can be re-expressed as 
\begin{eqnarray}
&&\rho _{m}(\partial _{t}+\mathbf{f}_{m}\cdot \nabla )f_{m}^{i}
-\sum_{j}\partial _{j}( \nu \rho _{m}\partial _{i}f_{m}^{\ j}+\nu \rho
_{m}\partial _{j}\ f_{m}^{\ i}) -\sum_j \partial _{j}((\nu \rho
_{m})\partial _{j}(\nu \partial _{i}\ln \rho _{m}))  \notag \\
&=&-\nabla ^{i}\left( -\frac{d}{d(1/\rho _{m})}\frac{\varepsilon }{\rho _{m}}%
\right) ,  \label{genera_cm}
\end{eqnarray}%
where $\rho _{m}$ is the solution of the continuity equation (\ref%
{Continuity2}), as before. Because of the asymmetric entry of $D$ and $%
\tilde{D}$ in the action, the derived equation violates the time reversal
symmetry.

\subsection{Navier-Stokes Equation}

\label{sec:ns}

\label{sec_ns}

The above scheme is applicable to derive the Navier-Stokes equation (NS). A
derivation of NS for incompressible fluids in SVM has already been worked
out in Refs. \cite{yasue2,if}. Recently, in Ref. \cite{kk}, the present
authors have derived the complete (compressible) NS starting from an action
different from the previous works. Then it is found that the difference
between $\mathbf{f}$ and $\widetilde{\mathbf{f}}$ plays a crucial role and
the velocity field in NS corresponds to $\mathbf{f}_{m}$. In below, we
reproduce this derivation within the framework developed here.

As is discussed in Ref. \cite{kk}, we consider that the potential term, $%
\varepsilon $ (internal energy density) depends not only on $\rho _{m}$ but
also on the other parameter $\hat{s}$ representing the internal
thermodynamical state of the fluid element. In thermal equilibrium, $\hat{s}$
is identified as the specific entropy. Now, we write the variation as 
\begin{equation}
I_{\lambda }=I-\lambda \int_{a}^{b}dt\int d^{3}\mathbf{R}\sum_{i}\rho _{m0}E%
\left[ \lambda \mathbf{g}\cdot \left( \tilde{D}D\mathbf{x}\ (\mathbf{R},t)+%
\frac{1}{\rho _{m}}\nabla P\right) +\frac{T}{m}\delta _{\lambda }\hat{s}%
\right] +O(\lambda ^{2}),
\end{equation}%
where $T$ and $P$ are defined in analogy to the usual thermodynamic relation
as 
\begin{eqnarray}
T &=& \left. \frac{\partial ( m \varepsilon /\rho _{m})}{\partial \hat{s}}
\right|_{\rho _{m}}, \\
P &=&-\left. \frac{\partial }{\partial (1/\rho _{m})}\left( \frac{%
\varepsilon }{\rho _{m}}\right) \right\vert _{\hat{s}}.
\end{eqnarray}%
In the thermodynamic equilibrium, these quantities coincide with the usual
temperature and pressure.

Because we are considering dissipative processes, the entropy of a fluid
element is not a conserved quantity. Then, to complete the variations, we
need information about $\delta _{\lambda }\hat{s}$ for the variation (\ref%
{vari}). In general, $\hat{s}$ is a scalar function of local quantities
(does not depend explicitly on $\mathbf{R}$) determined only by the
properties of the fluid element. Therefore, the most general form of $\hat{s}
$ depends only on the density $\rho _{m}$ and its derivatives$,\dot{\rho}%
_{m}\ $and$\ \nabla \rho _{m}$. Due to the vector nature of $\nabla \rho
_{m},$ this does contribute only to the higher order corrections of NS.
Thus, the most general form of $\hat{s}$ for our purpose is expressed as $%
\hat{s}=\hat{s}(\rho _{m},\dot{\rho}_{m})$.

Suppose that the hydrodynamic time scale $\tau _{hyd}\equiv \rho _{m}/\dot{%
\rho}_{m}$ is much larger than that of the microscopic degrees of freedom
inside a fluid element, $\tau _{mic}$. Note that the local thermal
equilibrium should be perfectly achieved for the ideal fluid where $\hat{s}$
is constant $(\delta _{\lambda }\hat{s}=0)$. Therefore, we expect that the
specific entropy is expressed in powers of $\tau _{mic}/\tau _{hyd}$ in such
a way that the ideal fluid case is recovered in the vanishing limit of $\tau
_{mic}$. We then write $\hat{s}=\hat{s}_{eq}+a_{1}\tau _{mic}\dot{\rho}%
_{m}/\rho _{m}+a_{2}(\tau _{mic}\dot{\rho}_{m}/\rho _{m})^{2}+\cdots $,
where $\hat{s}_{eq}$ is a constant for the variation and $a_{i}$'s are
expansion coefficients. Thus the lowest order truncation gives 
\begin{equation}
\delta _{\lambda }\hat{s}=\delta _{\lambda }\left( g_{\tau _{mic}}(\rho _{m})%
\dot{\rho}_{m}\right) ,  \label{delta_s}
\end{equation}%
where $g_{\tau _{mic}}(\rho _{m})$ is an arbitrary function of $\rho _{m}$.
For the stochastic variation, $\dot{\rho}_{m}$ is interpreted as $(D+\tilde{D%
})\rho _{m}/2$. Note that $\delta _{\lambda }\hat{s}$ is the virtual change
of $\hat{s}$ associated with the variations and does not necessarily satisfy
the thermodynamic principles such as $\delta _{\lambda }\hat{s}\geq 0$.

Then the condition of $\delta _{\lambda }I=0$ leads to 
\begin{equation}
\rho _{m}\left[ \frac{\partial }{\partial t}+\mathbf{f}_{m}\cdot \nabla %
\right] \mathbf{f}+\nabla (P-\mu \nabla \cdot \mathbf{f}_{m})-\left( \nabla
\cdot \eta \nabla \right) \mathbf{f}=0,  \label{general}
\end{equation}%
where $\eta $ is the shear viscosity. Here we used that $\nu =\eta /\rho_{m} 
$ and $\mu = - \rho_m^3 g(\rho _{m})/m ~(\partial T/\partial \rho_m)_{\hat{s}%
}$. One can see that the contribution from $\delta _{\lambda } \hat{s}$
effectively changes pressure by $\mu \nabla \cdot \mathbf{f}_{m}$.

Equation (\ref{general}) contains the two different velocity fields, $%
\mathbf{f}_{m}$ and $\mathbf{f}$, but the latter can be eliminated by using
Eq. (\ref{vel_ave}). As a result, we obtain 
\begin{eqnarray}
&&\rho \left[ \frac{\partial }{\partial t}+\mathbf{f}_{m}\cdot \nabla \right]
\mathbf{f}_{m}+\nabla (P-\zeta \nabla \cdot \mathbf{f}_{m})-\nabla \cdot
\left( \eta \ \widehat{e}^{m}\right)  \notag \\
&&-\left( \nabla \cdot \eta \nabla \right) \left( \frac{\eta }{\rho _{m}}%
\nabla \ln \rho _{m}\right) =0,  \label{NS}
\end{eqnarray}%
where $\widehat{e}^{m}=\left\{ e_{ij}^{m}\right\} $ is a $\left( 3\times
3\right) $ irreducible symmetric tensor, defined by 
\begin{equation}
e_{ij}^{m}=\partial _{j}\ f_{m}^{~i}+\partial _{i}\ f_{m}^{~j}-\frac{2}{3}%
(\nabla \cdot \mathbf{f}_{m})\delta _{ij},
\end{equation}%
and $\zeta =\mu +2\eta /3$. Identifying $\zeta $ as the bulk viscosity, Eq. (%
\ref{NS}) is nothing but the Navier-Stokes equation, except for the last
term on the left hand side.

The last term is not only of second order for the magnitude of fluctuations
(noise) $\nu =\eta /\rho _{m}$, but also of third order for the spatial
derivative $\nabla$. The Navier-Stokes equation does not contain such a term
since, by construction, only the first and second order spatial derivative
terms are maintained. In the case of rarefied gases, this corresponds to the
first order truncation in the derivative (Chapman-Enskog) expansion of
one-particle distribution functions.

On the other hand, in our case, the viscosity coefficients are directly
related to the size of noises. When the effect of fluctuations is large, we
cannot neglect the last term of Eq. (\ref{NS}). This term comes from the
difference between $\mathbf{f}$ and $\mathbf{f}_{m}$. Therefore, in such a
case, hydrodynamics is described by using the two velocities as Eq. (\ref%
{general}). This feature resembles that of the generalized hydrodynamics
proposed by Brenner \cite{brenner,other}.

\subsection{Gross-Pitaevskii Equation}

The Gross-Pitaevskii equation is a non-linear Schr\"{o}dinger equation and
has been used to study some aspects of the Bose-Einstein condensates \cite%
{GP}. The derivation of the Gross-Pitaevskii equation (GP) also has been
studied from various points of view \cite{GPeq}. In particular, it is shown
that GP can be derived in terms of SVM, extending the method of Sec. \ref%
{sec:ave} to a system of a many-body Lagrangian \cite{gross}. In this
subsection, we show that the formalism of SVM for a continuum medium can
readily be used to derive GP in a direct way.

We consider a system composed of identical bosons with mass $m$, under the
external trapping potential $V(\mathbf{r}).$ Supposing the interaction among
bosons is local, we write the potential energy density of the system as 
\begin{equation}
\varepsilon =V(\mathbf{r})\frac{\rho _{m}}{m}+\frac{1}{2}U_{0}\frac{\rho
_{m}^{2}}{m^{2}},
\end{equation}%
where $\rho _{m}=\rho _{m}\left( \mathbf{r},t\right) $ is the mass density
of the boson and $U_{0}$ represents the strength of interaction among them.
To obtain the equation for the wave function (order parameter), the kinetic
term of the action should be replaced by the average of $D\mathbf{x}$ and $%
\tilde{D}\mathbf{x}$ as was done in Sec. \ref{sec:ave}. Then, in the
Lagrangian coordinate system, we can define the action of SVM as 
\begin{eqnarray}
I &=&\int_{b}^{a}dt\int d^{3}\mathbf{R}\rho _{m0}(\mathbf{R})  \notag \\
&&\times \left[ \frac{1}{2}\frac{D\mathbf{x} \cdot D\mathbf{x}+ \tilde{D}%
\mathbf{x} \cdot \tilde{D}\mathbf{x}}{2}-\frac{1}{m}V(\mathbf{x})-\frac{1}{%
2m^{2}}U_{0}\rho _{m}(\mathbf{x},t)\right] .  \label{act_GP}
\end{eqnarray}%
Following the similar procedure in the previous subsection, SVM leads to 
\begin{equation}
(\partial _{t}+\mathbf{f}_{m}\cdot \nabla ) f^i_{m}-2\nu ^{2}\nabla^i (\rho
_{m}^{-1/2}\nabla ^{2}\sqrt{\rho _{m}})+\frac{1}{m}\nabla^i \left( V+\frac{%
U_{0}}{m}\rho _{m}\right) =0,  \label{ori_GP}
\end{equation}%
which should be solved together with the Fokker-Plank equation for $\rho
_{m} $, i.e., Eq. (\ref{Continuity2}). This equations is same as that
discussed in Ref. \cite{bern} and known as the quantum Bernoulli equation.

To convert Eq. (\ref{ori_GP}) into the corresponding Schr\"{o}dinger-type
equation, we follow the same procedure discussed in Sec. \ref{sec:ave},
introducing the wave function (see Eqs. (\ref{phase1}) and (\ref{Psi})) with
a modification of Eq.(\ref{Psi}) by 
\begin{equation}
\psi =\sqrt{\frac{\rho _{m}}{m}}e^{i\theta },  \label{Psi_GS}
\end{equation}%
because $\rho _{m}$ is the mass density. We again assumed that the flow is
irrotational. The equation corresponding to Eq.(\ref{sch_SVM}) becomes 
\begin{equation}
i\partial _{t}\psi =\left[ -\nu \nabla ^{2}+\frac{1}{2m\nu }(V+U_{0}|\psi
|^{2})\right] \psi ,
\end{equation}%
and setting $\nu =\hbar /(2m)$, we arrive at GP, 
\begin{equation}
i\hbar \partial _{t}\psi =\left[ -\frac{\hbar ^{2}}{2m}\nabla
^{2}+V+U_{0}|\psi |^{2}\right] \psi .  \label{GP}
\end{equation}%
As before, from the definition of $\psi ,$ $|\psi |^{2}=\rho _{m}/m$ gives
the particle distribution function.

Note that the classical limit of GP is easily obtained by taking the
vanishing limit of $\hbar $ (or $\nu $) in Eq.(\ref{ori_GP}), 
\begin{equation}
(\partial _{t}+\mathbf{f}_{m}\cdot \nabla ) f^i_{m}=-\frac{1}{m}\nabla^i V-%
\frac{U_{0}}{m^{2}}\nabla^i \rho _{m},
\end{equation}%
together with Eq. (\ref{Continuity2}). Of course, these equations are same
as those obtained from the action of Eq. (\ref{act_GP}) applying the
classical variational method.

The above argument is easily extended to include the external gauge fields
via the minimum coupling as described in Sec. \ref{sec_ele}. The required
action in the Lagrangian coordinate system is 
\begin{equation}
I=\int_{t_{a}}^{t_{b}}dt\int d^{3}\mathbf{R}\rho _{0}\left[ \frac{1}{2}%
\mathbf{v}^{2}+\frac{e}{cm}\mathbf{v}\cdot \mathbf{A}-\frac{1}{m}V(\mathbf{x}%
)-\frac{1}{2m^{2}}U_{0}\rho _{m}(\mathbf{x},t)-\frac{e}{m}\phi \right] ,
\end{equation}%
and the equation corresponding to Eq.(\ref{eqvm_ele}) becomes 
\begin{eqnarray}
(\partial _{t}+\mathbf{f}_{m}\cdot \nabla )\mathbf{f}_{m}-2\nu ^{2}\nabla
(\rho _{m}^{-1/2}\nabla ^{2}\sqrt{\rho _{m}})+\frac{1}{m}\nabla \left(
V+e\phi +\frac{1}{m}U_{0}\rho _{m}\right) &=&-\frac{e}{cm}\partial _{t}%
\mathbf{A}+\frac{e}{cm}\mathbf{f}_{m}\times (\nabla \times \mathbf{A}). 
\notag \\
&&
\end{eqnarray}%
The corresponding equation for the wave function is given by%
\begin{equation}
i\hbar \partial _{t}\psi =\left[ \frac{1}{2m}\left( -i\hbar \nabla -\frac{e}{%
c}\mathbf{A}\right) ^{2}+\left( V+e\phi +U_{0}|\psi |^{2}\right) \right]
\psi ,
\end{equation}%
where $\psi $ is given by Eq. (\ref{Psi_GS}) with the phase function given
by Eq. (\ref{phase2}), and the noise is set as $\nu =\hbar /(2m)$.

\subsection{Generalized Diffusion Equation and Physical Meaning of
Correction to Navier-Stokes Equation}

\label{sec:diff}

In Sec. \ref{sec:ns}, we found that NS derived using SVM involves the higher
order correction term, which was neglected for deriving NS. In this section,
we discuss the physical meaning of this correction term.

For the sake of simplicity, we consider the case where there is no potential
energy density in the action (\ref{ac_field}). Physically, this situation
will correspond to the free diffusion process. Then, from Eq. (\ref%
{genera_cm}), together with the continuity equation, we have, 
\begin{subequations}
\begin{eqnarray}
&&\partial _{t}\rho _{m}+\nabla \cdot (\rho _{m}\mathbf{f}_{m})=0, \\
&&\rho _{m}(\partial _{t}+\mathbf{f}_{m}\cdot \nabla )f_{m}^{\
i}-\sum_{j}\partial _{j}\{\nu \rho _{m}\partial _{i}f_{m}^{\ j}+\nu \rho
_{m}\partial _{j}\ f_{m}^{\ i}\}-\partial _{j}((\nu \rho _{m})\partial
_{j}(\nu \partial _{i}\ln \rho _{m}))=0.  \notag \\
&&  \label{GDE}
\end{eqnarray}
\end{subequations}
The first equation is the continuity equation as before, and the
second one is obtained by SVM, that is, Eq. (\ref{genera_cm}). In
the discussion of Sec.\ref{sec_ns}, the third term on the left hand side of
Eq. (\ref{GDE}) was neglected as the higher order term of NS, 
but now we will keep it in this subsection.

This equation describes the generalized diffusion process. Suppose that
there is a clear separation of time scales between $\rho _{m}$ and $\mathbf{f%
}_{m}$ in the \textit{Lagrangian} coordinate, and we observe the macroscopic
dynamics of $\rho _{m}$ where the material derivative of 
$\mathbf{f}_{m}$ is negligibly small. Then, from the second equation, we obtain 
\begin{equation}
\mathbf{f}_{m}=-\frac{\nu }{2}\nabla \ln \rho _{m}.
\end{equation}%
This corresponds to Fick's law of the diffusion equation. In fact, by
substituting it into the first equation, we obtain the usual diffusion
equation (see also Appendix \ref{app:a}), 
\begin{equation}
\partial _{t}\rho _{m}=\frac{\nu }{2}\nabla ^{2}\rho _{m}.  \label{nde}
\end{equation}

Differently from the diffusion equation, Eq. (\ref{GDE}) involves a memory
effect through the time derivative of $\mathbf{f}_{m}$. In this form, 
we see that Eq. (\ref{GDE}) is interpreted as the generalized
diffusion equation, and it reproduces the usual diffusion equation in the
Markov limit where the material derivative of $\mathbf{f}_{m}$ is neglected.

In other words, to reproduce the diffusion equation, we cannot neglect the
third term on the left hand side of the second equation, which corresponds
to the higher order correction term for NS. In this sense, this higher order
term may have a special meaning and be an important correction even in NS.

\section{Concluding Remarks}

In this paper, we discussed the application of the stochastic variational
method to particle systems and continuum mediums. The application to the
particle systems has been studied from various points of view. First, we
shortly summarized these results showing the derivation of the
diffusion-type equation and the Schr\"{o}dinger equation with and without
the minimal gauge coupling.

We further generalized of the idea of SVM to the calculations of continuum
mediums, and showed that the Navier-Stokes, Gross-Pitaevskii and generalized
diffusion equations are derived.

As is known, the usual diffusion equation is inconsistent with underlying
microscopic dynamics and should be modified \cite{km,koide_dif}. In fact, it
is shown that the coarse-grained equation obtained from microscopic
Hamiltonian using the projection operator method is not the usual diffusion
equation but the Maxwell-Cattaneo-Vernotte-type generalized diffusion
equation. This equation has several advantages compared to the usual
diffusion equation. For example, this generalized equation is consistent
with an exact relation obtained from microscopic dynamics, while the
diffusion equation is not \cite{koide_dif, koide_review}. Here, we proposed
a new generalized diffusion equation in the framework of SVM. It is
interesting to ask the relation between this generalized equation from SVM
and the Maxwell-Cattaneo-Vernotte-type equation.

We found that SVM predicts the higher order correction term to NS. This term
becomes important when there is strong inhomogeneity of the mass density. In
other words, we cannot recognize this effect from the study of
incompressible fluids. The importance of the correction term becomes more
clear in the discussion of the generalized diffusion equation. In order to
reproduce the usual diffusion equation in the Markov limit, the same higher
order term in the generalized diffusion equation plays an important role. In
this sense, this term may be an important correction even to NS.

There are several attempts to derive macroscopic equations for high density
matters by assuming a Langevin dynamics for underlying microscopic degrees of
freedom \cite{kawasakidean}. There, the interaction of the Langevin dynamics
is assumed in advance, while it is determined by applying the variational
principle in our approach. It is thus interesting to discuss the meaning of
the correction term which we found, comparing with the behavior of such a
dynamical density functional theory. More detailed study of this correction
term is a future task \footnote{There is an attempt to derive the dynamical density functional theory 
using the time-dependent projection operator method \cite{tdpom}. There,  
the so-called Q-approximation is employed. The problem of this approximation is discussed in Ref. \cite{koide_dif}. }.

The advantage of SVM is that dissipative equations can be derived
systematically once the action of the corresponding reversible dynamics is given. For
example, in the physics of soft matter, dissipative equations are usually
derived by applying Onsager's variational principle \cite{onsager,doi}. SVM may
serve another convenient approach to discuss the dynamics of such a complex
fluid.

So far, we have assumed that the noise of SVM is a Gaussian white noise. It
might be possible to generalize this discussion to the case of
multiplicative noises. For the present work, we assume that the property of
the noise is unchanged both for the forward and backward SDEs. This seems to
be reasonable assumption because the properties of the noise is completely
independent of the dynamics of macroscopic variables (Brownian particles) in
this case of the additive noise. This is, however, not trivial for the multiplicative noise. If
this assumption is still applicable, the extention of the present argument
to the multiplicative noise will be straightforward.

As was mentioned in the introduction, the essence of the classical
variational principle becomes more fundamental in the application of the
approach to field theories and relativistic systems. As far as we know, SVM
has not yet applied to such cases \cite{zast}. Studies of
relativistic heavy ion collisions revealed that it will be of the
fundamental importance to discuss dissipative mechanisms in relativistic
systems. Thus, an extension of the present approach to relativistic field
theoretical systems is expected to furnish an important clue to formulate
such processes from the fundamental concept of coarse-graining procedure to arrive
at the relativistic stochastic differential equation. 
%Then the effect of dissipation in high-energy physics can be studied systematically \cite{rot}.
However, the Lorentz covariant SDE still stays an
open question \cite{review_rbm}.

\hspace{1cm}

This work was financially supported by CNPq, FAPERJ, CAPES and PRONEX.

\appendix

\section{another limit to diffusion equation}

\label{app:a}

In Sec. \ref{sec:diff}, the generalized diffusion equation is derived and
the equation is reduced to the usual diffusion equation in the Markov limit.
Interestingly, this usual diffusion equation is still reproduced when a
special initial condition is chosen.

Suppose that, as the initial condition of $\mathbf{f}_m$, we have 
\begin{equation}
\mathbf{v}_m = -\nu \nabla \ln \rho.
\end{equation}
Then we can easily check that both of the first and second equations of Eq. (%
\ref{GDE}) are reduced to the same equation, 
\begin{equation}
\partial_t \rho_m = \nu \nabla^2 \rho_m .
\end{equation}
That is, the generalized diffusion equation is reduced to the usual
diffusion equation even when we use the special initial condition. The
difference from Eq. (\ref{nde}) is only the magnitude of the diffusion
coefficient.

\end{document}